\documentclass[aip,preprint,epsfig,showpacs]{revtex4-1}
\usepackage{subfigure}
\usepackage{graphicx}
\usepackage{epsfig}
\usepackage{amsmath}
\usepackage{xcolor}
\begin{document}
\title{Rupture of DNA Aptamer: new insights from simulations \\}
\author{Rakesh Kumar Mishra}
\author{Shesh Nath}
\author{Sanjay Kumar}
\affiliation{Department of Physics, Banaras Hindu University, Varanasi 221 005,India}

\begin{abstract}
Base-pockets (non-complementary base-pairs) in a double-stranded DNA
play a crucial role in biological processes. Because of thermal fluctuations, 
it can lower the stability of DNA, whereas, in case of DNA aptamer, small 
molecules e.g.  adenosinemonophosphate(AMP), adenosinetriphosphate(ATP) etc, form 
additional hydrogen bonds with base-pockets termed as 
``binding-pockets", which enhance the stability. Using the Langevin Dynamics 
simulations of coarse grained model of DNA followed by atomistic simulations, 
we investigated the influence of base-pocket and binding-pocket on the stability 
of DNA aptamer. 
Striking differences have been reported here for the separation 
induced by temperature and force, which require further investigation by single molecule experiments.
\end{abstract}
%\pacs{87.15.A-,64.70.qd,05.90.+m,82.37.Rs}
\maketitle

\section{Introduction}
Aptamers are Guanine(G)-rich short oligonucleic acids (DNA, RNA), which can perform 
specific function \cite{Gold, Szostak}. 
These have been developed {\it in vitro} through SELEX (Systematic 
Evolution of Ligands by Exponential Enrichment) process 
for  better understanding of the behavior of antibodies, which are produced
{\it in vivo} or in living cells \cite{Gold,Szostak,tan}. 
%They are Guanine (G)-rich short oligonucleic acids (DNA, RNA), which can perform specific function \cite{Gold, Szostak}.
One of the most extensively characterized 
examples of aptamer is found in telomerase at the ends of eukaryotic chromosomes, where it 
plays an important role in gene regulation \cite{sarkies,bejugam}. 
DNA loops or base-pockets consisting of Guanine can interact with small molecules
and proteins and thus enhance their stability, affinity and specificity \cite{Keefe}. 
For example, adenosinemonophosphate (AMP)  binds to DNA loop (termed as binding-pocket)
with eight hydrogen bonds, that increases the stability \cite{Lin, Teller,patel} of
aptamer. It served as the target of drugs for cancer treatment \cite{patel1}.
Their high affinity and selectivity with target 
proteins make them ideal and powerful probes in biosensors and potent 
pharmaceuticals \cite{Sullenger, Hoppe, Cho, chen, mao, Zayats, Zuo, Boyacioglu, Sun, Zhu, Lele}. 
Thus, understanding of the conformational stability of DNA aptamers before and after 
the release of drug molecule, and the influence of the binding of other molecules, are of crucial importance.

\begin{figure}[t]
\includegraphics[width=3.0in]{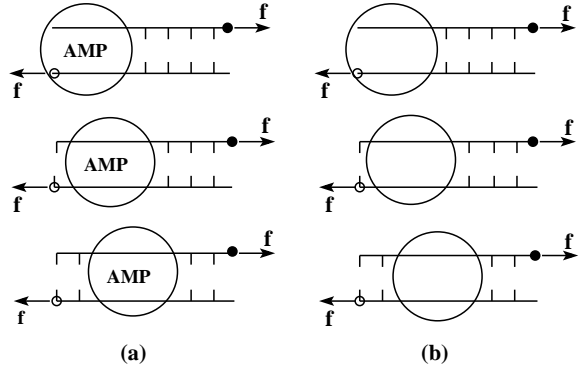}
\caption{Schematic representations of dsDNA under a shear force applied at
the opposite ends with varying position of (a) binding-pocket (AMP),
(b) base-pocket (without AMP). In the constant velocity simulation, one
end marked by blank circle is kept fixed, while in the constant force simulation,
the force is applied at both ends. The binding-pocket
(AMP within circle)Fig. 1(a) corresponds to additional binding of hydrogen bonds with AMP.
The circle in Fig. 1(b) represents the base-pocket
consisting of $G$ (say), where base-pairing is absent.
}
\label{fig-1}
\end{figure}
A double-stranded DNA (dsDNA) can be separated in two single-stranded DNA 
by increasing the temperature and the process is termed as DNA melting.
Traditional spectroscopic techniques used e.g. fluorescence spectroscopy, $UV-${\it vis} 
spectroscopy etc.  usually provide average responses for molecular interactions \cite{wartel}. 
Single molecule force spectroscopy (SMFS) techniques have emerged
as valuable tools for measuring the molecular interactions on a 
single molecule level, and thus unprecedented information about the stability of 
bio-molecules have been achieved. 
For example, DNA rupture  induced by SMFS techniques 
have been used to understand the strength of hydrogen bonds in nucleic acids, 
ligands-nucleic interaction, protein-DNA interactions etc. 
\cite{Lee_Science94,Strunge,hatch,cludia}.  In DNA rupture all intact base-pairs break 
simultaneously and two strands get separated, when a force is applied either at $3' -3'$ or $5' -5'$ ends. 
This force  is identified as rupture force.
Motivated by these studies, attempts have recently been 
made to understand the enhanced stability of DNA aptamers \cite{Nguyen, 
papa,Zlatanova,Yu,Jiang}. For example, Nguyen et al. \cite{Nguyen} measured 
the changes in rupture force of a DNA aptamer (that forms binding-pocket) 
with AMP (Fig. 1(a)) and without AMP (Fig. 1(b)), 
and thereby determined the dissociation constant at single-molecule level.
Papamichael et al. \cite{papa} used an aptamer-coated probe and an IgE-coated 
mica surface to identify specific binding areas. Efforts have also been made to 
determine the rupture force of aptamer binding to proteins and cells, and it was  
revealed that the binding-pocket enhances the rupture force by many 
folds \cite{Zlatanova,Yu,Jiang}.

Despite the progress made, there are noticeable lack of investigations.
For example, the melting of DNA aptamer in presence and absence of AMP and its dependence on the 
position in a DNA strand need to be measured and understood correctly. 
The aim of this paper is to develop a theoretical model to understand the role 
of a base-pocket and a binding-pocket on the rupture of a DNA aptamer. 
For this, a coarse grained model of DNA is developed.
Here, dsDNA is made up of two segments. One of the segments
consists of a DNA loop or the base-pocket of eight G type nucleotides. The other segment is stem,  
which is made up of twelve G-C base-pairs. We vary the position of the binding-pocket (Fig. 1 (a)) and the 
base-pocket (Fig. 1(b)) along the chain, and measure the rupture force and melting temperature 
for both cases (with and without AMP). For the first time, we report the profile as a function
of base/binding-pocket position. We find that even though  the melting profile has one 
minima (U-shape), there are two minima (W-shape) for rupture. This reflects that there are 
two symmetric positions, where the system can be more unstable. Extensive atomistic 
simulations have been performed to validate these findings, which helped us to 
delineate the correct understanding of the role of base-pockets in the stability of DNA aptamer.
In Sec. II, we briefly describe the model and the Langevin dynamics simulation to study the rupture of 
DNA aptamer \cite{mishra,cg}. The discussion on the melting and rupture profile of DNA aptamer as 
a function of base/binding-pocket position has also been made in this section. In Sec. III, 
we briefly explain the atomistic simulation \cite{Case, Duan,nath} and discuss the   
model independency of the results. Finally in Sec. IV, we conclude with a discussion on 
some future perspectives.
%\section{Coarse Grained Simulation}

\section{Model and Method}
We first adopt a minimal model introduced in Ref. \cite{mishra} for a homo-sequence of 
dsDNA consisting of $N$ base-pairs, where covalent bonds and base-pairing interactions are 
modelled by harmonic springs and Lennard-Jones (LJ) potentials, respectively. By 
using  Langevin dynamics simulation, it was shown that the rupture 
force and the melting temperature remain qualitatively similar to
the experiments \cite{Lee_Science94, Strunge,hatch,cludia}. 
Energy of the model system \cite{mishra,Allen,Smith} is given by.
\begin{eqnarray}
& & E   =  {\sum_{l=1}^2\sum_{j=1}^N}k({\bf r}_{j+1,j}^{(l)}-d_0)^2
+{\sum_{l=1}^2\sum_{i=1}^{N-2}\sum_{j>i+1}^N}4\left(\frac{C}{{{\bf r}_{i,j}^{(l)}}^{12}}\right) \nonumber \\
 & & + {\sum_{i=1}^N\sum_{j=1}^N}4\epsilon\left(\frac{C}{(|{\bf r}_i^{(1)}-{\bf r}_j^{(2)}|)^{12}}-
\frac{A}{(|{\bf r}_i^{(1)}-{\bf r}_j^{(2)}|)^6}\delta_{ij}\right),
\end{eqnarray}
where $N$ is the number of beads in each strand. Here, ${\bf r}_j^{(l)}$ represents the position of  bead $j$
on strand $l$. In the present case, $l=1(2)$ corresponds to first (complementary)
strand of  dsDNA. The distance between intra-strand beads, ${\bf r}_{i,j}^{(l)}$, is defined as
$|{\bf r}_i^{(l)}-{\bf r}_j^{(l)}|$. 
The harmonic (first) term with spring constant $k = 100$  couples the adjacent beads along each 
strand. The parameter $d_0 (=1.12)$ corresponds to the equilibrium distance in the harmonic 
potential, which is close to the equilibrium position of the LJ potential. The second term 
takes care of excluded volume effects {\it i.e.}, two beads can not occupy the same 
space \cite{scaling}. The third term, described by the Lennard-Jones (LJ) potential, takes care 
of the mutual interaction between the two strands. The  first term of LJ potential 
(same as second term of Eq.1) will not allow the overlap of two strands. 
The second term of the LJ potential corresponds to the base-pairing between
two strands. The base-pairing interaction is restricted to the native contacts
($\delta_{ij}=1$) only {\it i.e.},  the $i^{th}$ base of the $1^{st}$ strand forms pair 
with the $i^{th}$ base of the $2^{nd}$ strand only. It is to be noted here that $\epsilon$ 
represents the strength of the LJ potential. In Eq. 1, we use dimensionless distances 
and energy parameters and  set $\epsilon=1$, $C = 1$ and $A= 1$, which corresponds to a 
homosequence dsDNA. The  binding-pocket and base-pocket can be modelled by substituting 
$A = 1$  $ { \& }$ $ {\epsilon =2}$ (Fig. 1(a)) and $A=0$ ${ \& }$ $ { \epsilon =1}$ (Fig. 1(b)), 
respectively, among the bases inside the circle. The equation of motion is obtained 
from the following Langevin equation:

\begin{figure}[t]
\hspace{0.25 in}\includegraphics[width=3.in]{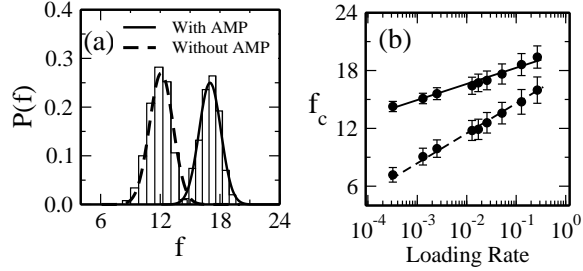}
\caption{(a) Probability distribution of rupture force of DNA aptamer with AMP(solid line) 
and without AMP(dashed line) for the same loading rate(0.0205). 
(b) Variation of rupture force with loading rate.}
\label{fig-2}
\end{figure}

\begin{equation}
m\frac{d^2{\bf r}}{dt^2} = -{\zeta}\frac{d{\bf r}}{dt}+{\bf F_c(t)}+\bf{\Gamma(t)},
\end{equation}
where $m ( =1 )$ and $\zeta (=0.4)$ are the mass of a bead and the friction
coefficient, respectively. Here, ${\bf F_c}$ is defined as $-\frac{dE}{d{\bf r}}$ and
the random force ${\bf \Gamma}$ is a white noise \cite{Smith},
i.e., $<{{\bf\Gamma}(t){\bf\Gamma}(t')}>= 6 \zeta T\delta(t-t')$. The $6^{th}$ order 
predictor-corrector algorithm with time step
$\delta t$=0.025 has been used to integrate the equation of motion.
The results are averaged over many trajectories.

First, we have calculated the rupture force of a dsDNA of two different
lengths \cite{text0}. We have inserted a base-pocket in the interior of the chain 
(circle in Fig. 1(b) and calculated the required force for 
the rupture at temperature $T = 0.12$. In order to obtain the rupture force 
for the DNA aptamer, we switched on the interaction among the base-pocket nucleotides with 
AMP (Fig. 1(a)): each of these extra interaction strength is double
of the base-pairing interactions. Following the experimental protocol \cite{Nguyen}, 
we performed  constant velocity  simulation \cite{Allen,Smith} by fixing one end of a DNA strand marked 
by a blank circle in Fig. 1. Force $f = K(vt - x)$ is applied on the other end of the strand 
marked by filled circle in Fig. 1. Here, $x$ is the displacement of the pulled monomer 
from its original position, $v$ is the velocity, $t$ is the time and $K (= 0.8)$ 
is the spring constant \cite{Li1}. The rupture force is identified as the maximum force, at which two strands separate suddenly. A selection of rupture force distribution of 500 events have been shown in Fig. 2(a).
For a given loading rate, the most probable rupture force $f_c$ is obtained by the Gaussian fit of the distribution 
of rupture force \cite{Strunge}. In Fig. 2(b), we have shown the variation of $f_c$ (with and without AMP) with 
the loading rate ($Kv$). It is apparent from the plots that the rupture force required 
for the  DNA aptamer is larger than the DNA (without AMP) with base-pocket, and 
the qualitative nature remains same as seen in the experiment \cite{Nguyen}.  

\begin{figure}[t]
\includegraphics[width=3.1in]{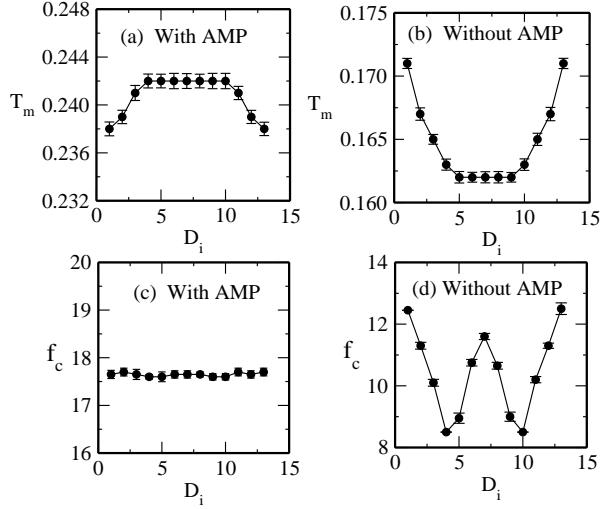}
\caption{(a) Variation of $T_m$ with binding-pocket positions
and (b) with base-pocket positions. (c) Variation of $f_c$ with
binding-pocket positions and (d) with base-pocket positions.}
\label{fig-3}
\end{figure}

For the better understanding of the role of base-pocket and
binding-pocket, we varied its position along the chain continuously 
from one end to the other, and calculated the melting temperature ($T_m$) and 
rupture force ($f_c$) for each position in the constant force ensemble (CFE) \cite{mishra}. 
The melting temperature is obtained here by monitoring the energy fluctuation ($\Delta E$) or the 
specific heat ($C$) with temperature, which are given by the following relations
\cite{Li}
\begin{eqnarray}
<\Delta E> & = & <E^2>-<E>^2  \\
C & = & \frac{<\Delta E>}{T^2}.
\end{eqnarray}
The peak in the specific heat curve gives the melting temperature.
 Since this simulation is in the equilibrium,  we have used  10 realizations with different seeds and
reported the mean value of the rupture force.
In Fig.3  (a)$-$(d), we have plotted the variation of $T_m$ (at $f=0$) and $f_c$ ($T=0.06$) 
much below $T_m$ as a function of pocket position ($D_i$). 
It is interesting to note that for thermal melting, the variation looks
like  $``\cap$ shape" for the DNA aptamer (with AMP) shown in Fig.3(a), and for without AMP, 
profile has ``U- shape" having one minima (Fig. 3(b). The rupture force remains invariant  
with binding-pocket positions for AMP (Fig.3(c)), which is consistent with experiment \cite{Nguyen}. 
Interestingly, in the absence of AMP, the profile has ``W-shape", {\it i.e.}, 
with two minima (Fig.3(d)). Although, one may intuitively expect the shape of profiles to be 
symmetric for the homosequence. But it is not apparent, why does the  profile has one 
minimum for the thermal melting and  two minima for the DNA rupture in the absence of AMP.

We now confine ourselves to understand these issues. In case of DNA aptamer melting, 
one would expect naively that the aptamer is more stable, when the binding-pocket is 
in the interior of the chain.  Thus, $T_m$ should be high (Fig. 3(a)) 
compare to the binding  pocket at the end resulting in a $``\cap$ shape" profile. In 
absence of AMP, the profile looks like ``U-shape".  
For a short chain, DNA melting is well described by the two state model \cite{sl} with  
$\Delta G= \Delta H - T \Delta S$, where $H$ and $S$ are the enthalapy and entropy, respectively.
The melting temperature $ T_m = \frac{\Delta H}{\Delta S}$ corresponds to the state with $\Delta G =0$
indicating that the system goes from the bound-state to the open-state and  change in the free energy of the 
system is zero. It is often practically easier to identify $T_m$ as the temperature where $50\%$ hydrogen 
bonds are broken. 

For a homo-sequence chain (without base-pocket), it was shown that the chain opens from the end 
rather than the interior of the chain \cite{shikha_jcp}. In such a case, the major contribution 
to the entropy comes from the opening of base-pairs near the end of the chain, and 
decreases to zero, when one approaches to the interior of the chain. 
In this case, there are two contributions to the entropy: entropy associated with opening of the
end base-pairs ($S_E$) and the base-pocket entropy ($S_{BP}$).
When a base-pocket is inserted at the end of DNA chain, there is an additional contribution 
to the  entropy, which is greater than $\Delta S_E$ or ($\Delta S_{BP}$), but less than the sum of two. Thus, $T_m$ decreases to $\sim 0.171$ from $\sim 0.181$ \cite{text0}. As the base-pocket moves along 
the chain, entropy of interior base-pairs increases. Combined effect of both entropies reduces the 
melting temperature further. Once the base-pocket is deep inside, the total entropy of the system
becomes equal to the sum of end-entropy and base-pocket entropy. As a result,
$T_m$ remains constant($\sim 0.161$) (Fig. 3(b)) irrespective of the base-pocket position. As base-pocket
approaches towards the other end, by symmetry we observe ``U-"shape profile. The conversion of reduced unit to real unit indicates that it is possible to observe it {\it in vitro} \cite{text}.

Understanding the decrease in $T_m$ with base-pocket position does not
explain, why the rupture profile of DNA has two minima.  It is clear 
from the Fig. 3 (b) \& (d) that the base-pocket entropy alone is not responsible for this ($W$)
shape. Based on the ladder model of DNA (homo-sequence), de Gennes proposed that the rupture force
$f_c$ is equal to $2 f_1 (\chi^{-1} \tanh(\chi \frac{N}{2}))$ \cite{degennes}.
Here, $f_1 $ is the force required to separate a single base-pair and  $\chi^{-1}$ is
the de Gennes length, which is  defined as $\sqrt{\frac{Q}{2R}}$. $Q$ and $R$ are the 
spring constants characteristic of stretching of the backbone and hydrogen bonds 
\cite{degennes}, respectively. 
de Gennes length is the length over which differential force is distributed. 
Above this length, the differential force approaches to
zero and there is no extension in hydrogen bonds due to the applied force \cite{hatch,mishra}.
When the base-pocket is at the end, four bases of one strand is under the tension,
whereas the complementary four bases are free, thereby increasing the entropy of the system.
The four bases act like a tethered length and a force is required to keep it
stretched so that rupture can take place. It is nearly equal to the ruptured force of   
a 12 base-pair dsDNA with no defect. As the base-pocket moves towards the center, the entropy of
base-pocket decreases as the ends of the base-pocket is now not free. 
Since, the de Gennes length for the base-pocket is infinite ($R =0$), it
implies that the differential force remains constant inside the base-pocket, thereby decreases
the stability of DNA, and  as a result, rupture force decreases further. The force 
acting at ends penetrates only up to the de Gennes length and within this, the rupture 
force keeps on decreasing as base-pocket moves and approaches to its minimum. Above a 
certain length, the rupture force starts increasing, which can be seen in Fig.3 (d) 
and approaches to the maximum at the middle of the chain. By symmetry, we get profile of two minima of ``W-shape".

\begin{figure}[t]
\includegraphics[width=1.7in]{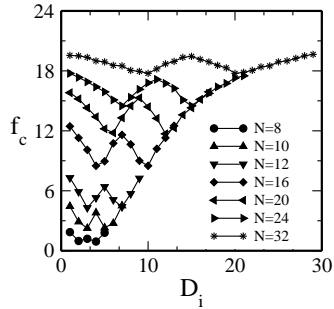}
\caption{ Variation of $f_c$ with base-pocket positions for different 
lengths.
}
\label{fig-4}
\end{figure}

It should be pointed here, that the de Gennes length in the present model is about ten
\cite{text3}, but for a DNA of length $16$ base-pairs (12 complementary and 4 non complementary), 
the minima is around $D_i = 4$. Note that the penetration depth for the 
two ends (say $5'-5'$) is different because of the asymmetry arising due to the presence of base-pocket, and 
hence the minima shifts. If we increase the length of DNA, keeping the base-pocket size constant, 
one would then expect that the minima will shift towards 10 (above the de Gennes length, 
the end-effect vanishes). This indeed we see in Fig. 4, where  the variation of $f_c$ 
with base-pocket position for different chain lengths ($N = 8,10,12,16,20,24$ and $32$)
has been plotted. In all cases, profiles have two minima, whereas minima shifts towards 
10 as $N$ increases. 
de Gennes equation predicts that $f_c$ increases linearly with length for 
small values of $N$, and saturates at the higher values of $N$, which is consistent 
with recent experiment \cite{hatch} and simulations \cite{mishra}. 
Surprisingly, right side of the profile also appears to follow the de Gennes equation.
This implies that, the base-pocket can reduce
the effective length of the chain, so that the rupture force is less compare
to the bulk value and approaches to a minimum value at a certain position. 
As the end-effect decreases, $f_c$ starts increasing and 
approaches to its bulk value. 

\section{Atomistic Simulation} 
In order to rule out the possibility that the above effect may be a consequence of 
the adopted  model,  we performed the atomistic simulations with explicit solvent. 
Here, we have taken homo-sequence DNA consisting of 12 G-C bps and a varying base-pocket of 
$8$  G-nucleotides \cite{Case,Duan,nath,Santosh}. The starting structure of the DNA 
duplex sequence having the base-pocket is built using make-na server\cite{ser}.
We used  AMBER10 software package \cite{Case} with all atom (ff99SB)
force field \cite{Duan} to carry the simulation. Using the LEaP module in AMBER, 
we add the AMP molecule (Fig. 5(a))  in the base-pocket and then   the $Na^{+}$ (counterions) to
neutralize the negative charges on phosphate backbone group of DNA structure (Fig. 5(b)). This
neutralized DNA aptamer structure is immersed in water box using TIP3P model for water \cite{Jorgensen}. We have
chosen the box dimension in such a way that the ruptured DNA aptamer structure remains fully inside the water
box. We have taken the box size of $57 \times 56 \times 183$ $\AA^3$
which contains $15674$ water molecules and $30$  $Na^{+}$ (counterions). 
A force routine
has been added in  AMBER10 to do simulation at constant force \cite{Santosh,text00}.
In this case, the force has been applied at
$5'-5'$ ends . The electrostatic interactions have been  calculated
with Particle Mesh Ewald (PME) method \cite{darden,essmann} using a cubic B-spline interpolation of
order 4 and a $10^{-5}$ tolerance is set for the direct space sum cut off. A real space cut off of
10 $\AA$ is used for both the van der Waal and the electrostatic interactions.
The system is
equilibrated at $F =0 $ for 100 ps under a protocol described in Ref. \cite{maiti,nanolett} and it 
has been ensured that AMP has bound with the base-pocket.
We carried out  simulations in the isothermal-isobaric (NPT) ensemble using a time step of
1 fs for 10 different realizations. We maintain the constant pressure by isotropic position scaling \cite{Case} with a
reference pressure of 1 atm and a relaxation time of 2 ps.  The constant temperature was
maintained at 300 K using Langevin thermostat with a collision frequency of 1 ps$^{-1}$. 
We have used $3D$  periodic boundary conditions  during the simulation.

\begin{figure}[t]
\includegraphics[width=3.0in, height=1.8in]{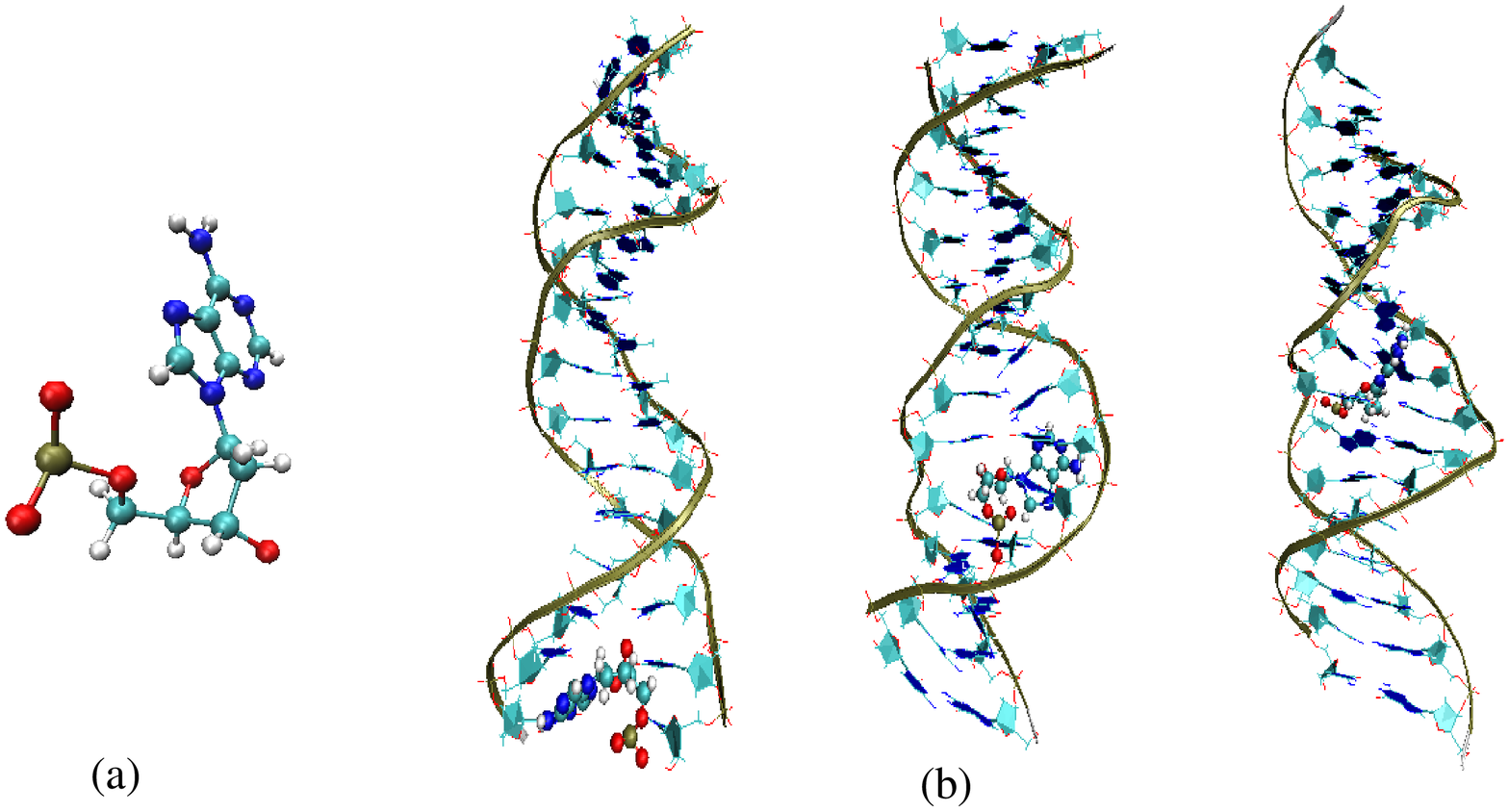}
\caption{(Color online) (a) Structure of AMP molecule, which has been inserted in the base-pocket.
(b) Snapshots of binding of AMP molecule at three  different positions ($D_i$=1, 4 and 7) 
in the DNA.}
\label{fig-5}
\end{figure}

Because of extensive time involved in the
computation, we restricted ourselves at three different (extremum) base-pocket positions 
$D_i$=1, 4 and 7  (Fig. 5(b)) and calculated the rupture force with 10 realization of different seeds
as a mean force.
%the peak value of the Gaussian distribution \cite{nath1}. 
The required rupture forces for these positions are 840 pN $\pm$ 20 pN, 720 pN  $\pm$  20 pN 
and 830 pN $\pm$ 20 pN, indicating that the complete profile contains two minima \cite{text2}.  
In presence of AMP, which interacts with the base-pocket, we find $f_c = $925pN $\pm$ 20pN 
remains constant irrespective of binding-pocket position as seen 
earlier (Fig. 3(c)). This validates the finding of simple coarse grained model, which captured 
some essential but unexplored aspects of rupture mechanism of DNA aptamer.

\section{Conclusions} 
Our numerical studies clearly demonstrate that the force induced rupture
and thermal melting of DNA aptamer will vary quite significantly.  In melting, all the nucleotides get almost 
equal thermal knock from the solvent molecules, whereas in DNA rupture, 
force is applied at the ends and the differential force acts only up to the 
de Gennes length \cite{degennes}. As a result, the stability of DNA in presence 
of base-pocket is strikingly different (single minima {\it vs} double minima). 
This may have biological/pharmaceutical significance because after the release of 
drug molecule, the stability of carrier DNA depends on the position from which it
is released. Hence, at this stage our studies warrant further investigations most likely 
by new experiments to explore the role of base-pockets and its position,  which will significantly 
enhance our understanding  about the stability of DNA aptamer and its suitability in the 
development and designing of drugs.

\section{Acknowledgements}
We thank G. Mishra, D. Giri and D. Dhar for many helpful discussions on the subject. 
We acknowledge financial supports from the DST, UGC and CSIR, New Delhi, India.

%\begin{thebibliography}{99}


\begin{thebibliography}{1}
\bibitem{cg} A FORTRAN code is developed to study the dynamics of the system.
\bibitem{text0} For a chain of 16 base-pairs, the rupture force ($f_c$) and the melting 
temperature ($T_m$) are $\sim 16$ and $\sim 0.181$ respectively. Whereas for 12 base-pairs, their values are $\sim 12$ and $\sim 0.171$, 
respectively.
\bibitem{text} Following relations may be used to convert dimensionless units to real units: 
$ T= \frac{{k_{b}}T^{*}}{\epsilon} $, $ t^{*} = (\frac{m\sigma^2}{\epsilon})^{\frac{1}{2}} t $ and 
$ r = \frac{r*}{\sigma} $ \cite{Allen}, where  $ T^{*}$, $ t^{*}, r^{*}$, and $ \epsilon $ are temperature, time, distance, and characteristic hydrogen bond energy in real units, respectively.  
$\sigma$ is the inter-particle distance at which the potential approaches  to zero. Now, if one sets effective base paring energy $ \epsilon \sim 0.17 $ $ ev $, which is obtained by comparing the melting temperature in the reduced $T_{m}$($ = 0.161$) to the one obtained from the oligo-calculator \cite{Kibbe} for the same sequence, one gets  $T_{m} \approx 56^{\circ}C$. By
setting average mass of each bead $ \approx 5 \times 10^{-22} g $ and $\sigma = 5.17 \AA$, one can obtained 
relation between real time $ t^{*} $ and reduced time $ t $  as $ t^{*} \sim 3 t $ ps.
\bibitem{text3} This approximate value can be obtained by putting different elastic constants involved in 
the expression of $\chi^{-1}$. In the present simulation, we have used $k = Q$ (=100) 
i.e. the spring constant associated with the covalent bonds along the backbone of the chain.
The spring constant $R$ associated with the base-pairs has been obtained by expanding  
the LJ potential  around its equilibrium value and setting it equal to zero above the
cut off distance, where rupture takes place. The second (harmonic) term of the expansion 
gives the elastic constant $R \sim 0.5$ associated with the LJ potential involved
in the base-pairing.
\bibitem{text2} Here, the simulation time is about $7ns$, and hence rupture force is found 
to be one order higher than AFM.
\end{thebibliography}
\end{document}